\documentclass[aps,prl,english,twocolumn,showpacs,preprintnumbers,amsmath,amssymb,floatfix,superscriptaddress,longbibliography]{revtex4-1}

\usepackage[T1]{fontenc}
\usepackage[latin9]{inputenc}
\usepackage{graphicx}
\usepackage{amssymb}
\usepackage{babel}
\makeatletter

\usepackage[version=3]{mhchem}

\usepackage{color}

\global\arraycolsep=2pt
\usepackage{stmaryrd}
\usepackage{amsmath}
\usepackage{amssymb}
\usepackage{graphicx}
\usepackage{textcomp}
\usepackage{calrsfs}
\usepackage{yfonts}
\usepackage{bm}
\usepackage{color}
\newcommand{\ben}{\begin{equation*}}
\newcommand{\een}{\end{equation*}}
\newcommand{\bean}{\begin{eqnarray*}}
\newcommand{\eean}{\end{eqnarray*}}

\newcommand{\nn}{\nonumber}
\newcommand{\be}{\begin{equation}}
\newcommand{\ee}{\end{equation}}
\newcommand{\bea}{\begin{eqnarray}}
\newcommand{\eea}{\end{eqnarray}}

\newcommand{\psumbar}{\sum\nolimits^\prime}
\DeclareMathOperator*{\psum}{\psumbar}

\makeatother
\usepackage{babel}
\makeatother

\usepackage{color}

\begin{document}

\title{Different pathways to anomalous stabilization of ice layers on methane hydrates}

\author{Y. Li}
 \email{leon@ncu.edu.cn}
  \affiliation{School of Physics and Materials Science, Nanchang University, Nanchang 330031, China}
  \affiliation{Institute of Space Science and Technology, Nanchang University, Nanchang 330031, China}

\author{I. Brevik}
\affiliation{Department of Energy and Process Engineering, Norwegian University of Science and Technology, NO-7491 Trondheim, Norway}

\author{O. I. Malyi}
\affiliation{Centre of Excellence ENSEMBLE3 Sp. z o. o., Wolczynska Str. 133, 01-919, Warsaw, Poland}

\author{M.  Bostr{\"o}m}
\email{mathias.bostrom@ensemble3.eu}
\affiliation{Centre of Excellence ENSEMBLE3 Sp. z o. o., Wolczynska Str. 133, 01-919, Warsaw, Poland}

\begin{abstract}
{We explore the Casimir-Lifshitz free energy theory for surface freezing of methane gas hydrates near the freezing point of water. The theory enables us to explore different pathways, resulting in anomalous (stabilising) ice layers on methane hydrate surfaces via energy minimization. Notably, we will contrast the gas hydrate material properties, under which thin ice films can form in water vapor, with those required in the presence of liquid water. It is predicted that methane hydrates in water vapor near the freezing point of water nucleate ice films but not water films.}
\end{abstract}
\date{\today}

\maketitle


\par The work of Elbaum and Schick\,\cite{Elbaum} on the Casimir-Lifshitz (CL) interaction in ice/water systems has   piqued the interest of scientists\,\cite{Dash1995,Wettlaufer,benet16,benet19,li2019interfacial,Prachi2019role}, as it sheds light on the formation of ice layers on gas hydrates, which are found  in ice-cold water~\cite{doi:10.1029/94GL01858}, ocean seabed~\cite{doi:10.1002/2016GL068656}, sediments~\cite{doi:10.1029/2007GC001920}, and in permafrost~\cite{NaturalGas2003}. With reliable, experimentally derived dielectric properties of ice and water\,\cite{JohannesWater2019}, authors initiated the theoretical work that predicts relatively thick ice layers on partially degassed hydrates, demonstrating a previously unconsidered mechanism for the stabilization of gas hydrates in cold water\,\cite{BostromEstesoFiedlerBrevikBuhmannPerssonCarreteroParsonsCorkery2021}.


\par As an experimentally well-known effect\,\citep{Handa1986,HallbruckerMayer1990,ErshovYakushev1992,TakeyaRipmeester2008,Falenty2009,HachikuboTakeyaChuvilinIstomin2011}, termed `self-preservation', it was previously assumed that ice layers were created through kinetics, causing gas molecules to deplete in the gas hydrate surface region and the surface molecular structure to reconstruct into an ice-like one. 
Our former works, as well as the observation of stable shallow hydrates~\cite{Shakhova2017,ChuvilinBukhanovDavletshinaGrebenkinIstomin2018,Shakhova2019}, suggest that under favorable conditions, the formation of the ice layer on the hydrate is not driven purely by kinetics, but also by influences on thermal equilibrium from the CL interaction. Surface tension effects can, to some extent, be assumed constant in quasi-planar geometry as the ice layer thickness varies, motivating that the CL stress can contribute non-trivially. Particularly, the formation of ice layers on methane hydrates in the cold water, arising from the CL interaction, was forecasted, and estimated to be almost micron-sized, or at least much more than several tens of nanometers. For micron-sized ice layers, the thermal fluctuation has impacts on the growth of ice layers. 
These findings, along with the well-documented observation of the self-preservation of hydrates, suggest that the CL interaction could play an important role in understanding self-preservation in gas transport technology, geotechnical and global environmental risk assessments, and other relevant contexts.
If the ice-coating on hydrates is indeed spontaneous, resulting in lowering Gibbs free energy as we find, it adds weight to the argument that gas hydrates may be stabilized in a mechanism not previously understood, highlighting the necessity to fully understand the role of CL interaction in the formation and self-preservation of hydrates.

\par In this letter, we concisely describe the theory and predict the phenomenon, stemming from the CL interaction, that different external conditions can bring about different pathways to self-preservation. We define a wet pathway as the case, in which the outer surface of the gas hydrate is in  contact with a bulk water reservoir. In contrast, for the dry pathway, this outer surface is in contact with water vapor. As shown below, some gas hydrates in water gain the ice-coating interface, whereas others have a wet surface.
The surface properties of gas hydrates, required to achieve ice layer formation in water, deviate from those in cold water vapor. 
We revisit the \emph{wet pathway}, through which ice layers form on hydrates immersed in a bulk of ice-cold water, and propose a new \emph{dry pathway}, where ice layers form between hydrates and the water vapor. The dry pathway leads to thick ice layers, specifically a layered methane hydrate-ice-water-vapor configuration.
The dry pathway is investigated with a new inhomogeneous CL theory\,\cite{ParasharQEstress2018,Esteso4layerPCCP2020,LUENGOMARQUEZMacDowell2021,LiMiltonBrevikMalyiThiyamPerssonParsonsBostrom_PRB2022}. The water layer in the dry pathway appears at the ice-vapor interface via the structural force\,\cite{LuengoMarquez_IzquierdoRuiz_MacDowell2022}. The required surface properties to establish anomalous ice layer follow from the classical arguments on Casimir-Lifshitz free energy, where the intermediate layer grows when it has dielectric properties in between the surrounding media\,\cite{Dzya}. Therefore, ice formation through the wet pathway requires a partly gas-depleted surface region, while the dry pathway, in contrast, needs a sufficiently high gas occupancy in the surface region of gas hydrate.
\begin{figure}
  \centering
  \includegraphics[width=1.0\columnwidth]{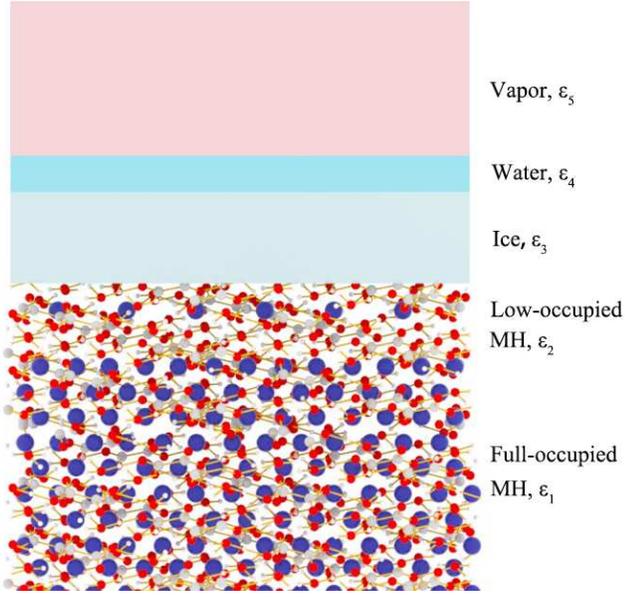}
  \caption{\label{Fig.SchematicFig1} (Color online) Schematic diagram of the five-layer system \ce{CH4} hydrate system: high-occupancy (HO) gas  hydrate $\varepsilon_1$; low-occupancy (LO) hydrate, $\varepsilon_2$; pure \ce{H2O} ice, $\varepsilon_3$; pure liquid \ce{H2O}, $\varepsilon_4$, and vapor, $\varepsilon_5$. In the wet pathway, the water layer is assumed to first go to infinite thickness (i.e. we consider a bulk water reservoir). In the dry pathway, the outer surface is in contact with water vapor. }
\end{figure}

\par We address the Casimir-Lifshitz free energy at finite temperature by utilizing recently reported dielectric functions for ice and water obtained by Luengo-M{\'a}rquez and MacDowell~\cite{LUENGOMARQUEZMacDowell2021}.
The quadruple point of methane hydrate (MH), very close to the triple point of water, is at about $T=272.9\, \rm{K},\ p = 25.63\,\rm{bar}$~\cite{Sloan}.
As in Refs.~\cite{Bostrom_acsearthspacechem.9b00019,BostromEstesoFiedlerBrevikBuhmannPerssonCarreteroParsonsCorkery2021}, we model the dielectric function of MH ($\varepsilon_{mh}$) using a Lorentz-Lorenz model~\cite{Aspnes} with mixing scheme for each gas hydrates from Bonnefoy et al.~\cite{BONNEFOY2005336,Bonnefoy}.
The important factors for the dielectric function of MH include the effective ice permittivity weighted by the density of water molecules in hydrates relative to pure ice, and the polarizability of methane molecules given in Ref~\cite{Johannes} and weighted with the gas density. (For more details, please see the Supplemental Material.)
We consider MHs covered with an ice layer that is sufficiently wide to be treated as locally planar. A schematic illustration of our model system is shown in Fig.\,\ref{Fig.SchematicFig1}. The CL free energy per unit area at temperature $T$, for a five-layer configuration, reads as~\citep{Ellingsen_2007,Esteso4layerPCCP2020,LiCorkeryCarreteroBerlandEstesoFiedlerMiltonBrevikBostrom2023}
\begin{eqnarray}
&F_{\rm CL}(d)&=\frac{T}{2 \pi}
\psum_{m=0}^\infty \int_0^\infty dkk\sum_{s=\rm E,H}\ln
\bigg[
1+r^{s}_{12}r^s_{23}e^{-2\kappa_2d_2}\nn\\&&
+r^s_{23}r^s_{34}e^{-2\kappa_3d_3}
+r^s_{34}r^s_{45}e^{-2\kappa_4d_4}\nn\\&&
+r^s_{12}r^s_{34}e^{-2(\kappa_2d_2+\kappa_3d_3)}
+r^s_{23}r^s_{45}e^{-2(\kappa_3d_3+\kappa_4d_4)}\nn\\&&
+r^s_{12}r^s_{23}r^s_{34}r^s_{45}e^{-2(\kappa_2d_2+\kappa_4d_4)}\nn\\&&
+r^s_{12}r^s_{45}e^{-2(\kappa_2d_2+\kappa_3d_3+\kappa_4d_4)}
\bigg],\label{100}
\end{eqnarray}
where $k$ is the modulus of the wave vector ${\bf k}$ parallel to the surface, $\kappa_i=\sqrt{k^2+\varepsilon_i\xi_m^2}$ with $i=1,\cdots,5$ labeling layers in the five-layer system with the layer-stacking order $1-2-3-4-5$, $\xi_m=2\pi mT$ being the $m$th Matsubara frequency, and $d_i$ being the thickness of the $i$th layer. The primed sum means the $0$th term ($m=0$) is weighted in half. (The natural unit $\hbar=c=\varepsilon_0=\mu_0=k_{\rm B}=1$ is employed, unless specified otherwise.) The $s=\rm{E}$ and $s=\rm{H}$ signify the contributions from the transverse electric (TE) and transverse magnetic (TM) modes, respectively. The reflection coefficients at the interface between $i$th and $j$th layers are defined as~\citep{Ellingsen_2007}
\begin{equation}
    r^{\rm E}_{ij} = \frac{\kappa_i-\kappa_j}{\kappa_i+\kappa_j},\qquad r^{\rm H}_{ij} = \frac{\varepsilon_j\kappa_i-\varepsilon_i\kappa_j}{\varepsilon_j\kappa_i+\varepsilon_i\kappa_j}.\label{eq:rtTETM}
\end{equation}
The CL free energy for the four- and three-layer systems can be obtained with Eq.~\eqref{100}, by setting the materials of properly chosen adjacent layers the same\,\cite{Esteso4layerPCCP2020,LUENGOMARQUEZMacDowell2021,LiCorkeryCarreteroBerlandEstesoFiedlerMiltonBrevikBostrom2023}. When accounting for structural forces  related to
the packing of water molecules on the solid ice substrate\,\cite{LuengoMarquez_IzquierdoRuiz_MacDowell2022,Karim1988TheII}, an atomically thin water layer arises on the interface between the anomalous-stabilizing ice layer and the vapor, which has no effect on the anomalous stabilization in this research, i.e. on how effective the ice layer prevents the leakage of methane molecules.


\section*{Results and Discussions}
\label{RD}

\par
\begin{figure}
  \centering
  \includegraphics[width=1.0\columnwidth]{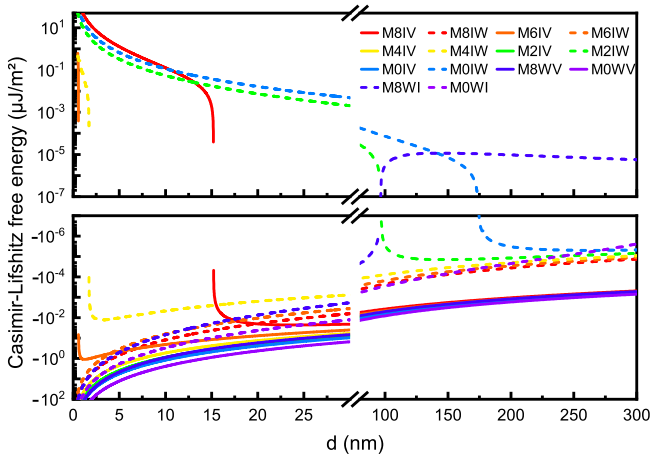}
  \caption{\label{Fig.3LF} (Color online) The Casimir-Lifshitz free energy per unit area for three-layer configurations MH ($n_{\rm b}=i$)-ice-vapor (M$i$IV), MH ($n_{\rm b}=i$)-ice-water (M$i$IW), MH ($n_{\rm b}=i$)-water-ice (M$i$WI), and MH ($n_{\rm b}=i$)-water-vapor (M$i$WV) with $n_{\rm b}$ being the occupancy number.}
\end{figure}

 For the bulk MH immersed in the reservoir of ice-cold water, as in Refs.~\cite{Bostrom_acsearthspacechem.9b00019,BostromEstesoFiedlerBrevikBuhmannPerssonCarreteroParsonsCorkery2021}, the growth of ice layer, due to the CL interaction, are seen. According to Fig.~\ref{Fig.3LF}, with a low occupancy ($n_{\rm b}<5$), the ice layer forms via a wet pathway and can have a thickness ranging from several to hundreds of nanometers, as this minimizes the CL free energy.
However, in contrast, when the MH is relatively fully occupied, the three-layer free energy can only be minimized by reducing the thickness of the ice layer to zero, preventing ice formation, or resulting in a wet MH surface. Furthermore, when the MH is in contact with a bulk of ice rather than water, the CL free energy, thus stress, prohibits the formation of even micron- or nano-sized water layers, resulting in stress.

\par When the MH is exposed to a water vapor environment, the ice layer should exist as demonstrated in Fig.~\ref{Fig.3LF}, if the MH is quite fully filled ($n_{\rm b}>5$), in contrast to the wet path scenario above. That is, this ice layer is formed via the dry pathway.
As expected and shown in Fig.~\ref{Fig.3LF}, a water layer does not present between the MH of any occupancy and the vapor, implying a dry MH surface. Remarkably, if the surface layer is degassed before the anomalous ice coating is formed, the ice can not form via the dry path, but must rely on the traditional mechanism\,\citep{Handa1986,HallbruckerMayer1990,ErshovYakushev1992,TakeyaRipmeester2008,Falenty2009,HachikuboTakeyaChuvilinIstomin2011} via the reorganization of surface layers.

\par The results above have not yet taken into account the possibility that both ice and water layers can appear on the surface of MH. Furthermore, a bulk of MH usually holds a surface region with lower occupancy, as a result of the rather slow degassing process. Cases like that entail the CL free energy in configurations containing more than three layers, where influences of the thickness of each layer should be examined explicitly. In the four-layer configuration with the media labeled by $i=1,2,3,4$ and stacked in the order $1$-$2$-$3$-$4$, the CL free energy $F_{\rm CL}$ can be divided into three parts, namely the three-layer contributions $F_{13}$ and $F_{24}$ from $1$-$2$-$3$ and $2$-$3$-$4$ respectively, and the pure four-layer contribution $F_{14}$,
\begin{eqnarray}
F_{\rm tot}(d_2,d_3)=F_{13}(d_2)+F_{24}(d_3)+F_{14}(d_2,d_3), \label{eqFtot}
\end{eqnarray}
where $d_i,\ i=2,3$ is the thickness of the $i$th layer.
If the bulk MH in a vapor environment has no degassed surface region, then any four-layer configuration, whether it involves MH-water-ice-vapor or MH-ice-water-vapor, will ultimately converge to the stable state predicted by the corresponding three-layer cases. The additional four-layer contribution is not significant enough to introduce other equilibrium. On the other hand, if a degassed region exists on the surface of bulk MH (the bulk MH is assumed fully filled for simplicity), then the thickness of the surface MH layer can be taken as a constant in the ice-forming process, since this surface region changes slowly. The CL free energy counted, therefore, should be the reduced one, obtained by omitting $F_{13}(d_2)$ in the configuration bulk MH-surface MH-X (water or ice)-vapor. The water layer between the surface MH and the vapor is always suppressed as in the three-layer cases in Fig.~\ref{Fig.3LF}. Similar results apply to the ice layer cases. Moreover, even when the MH material is immersed in cold water or covered by ice, the pure four-layer interaction is not enough to significantly modify the stabilities predicted by three-layer cases.

\subsection*{Discussion on different pathways to self-preservation}
\par It is now widely accepted that anomalous stabilization of ice layers on gas hydrate surfaces can form via kinetics, leading to the reconstruction of the outer layers of the hydrate. However, we have demonstrated that  ice layers relevant for anomalous stabilization of methane gas hydrates can also form via equilibrium thermodynamics mediated by the Casimir-Lifshitz force. Previous works\,\cite{BostromEstesoFiedlerBrevikBuhmannPerssonCarreteroParsonsCorkery2021} indicated that a surface region of the gas hydrate with gas molecule depletion was essential. However, this very much depends on the surrounding media. The low occupancy surface required in liquid water has been contrasted with a high occupancy surface region essential for ice formation in water vapor. Notably, our new theory accounts for two regions (liquid water and ice) growing and receeding at each other's expense. Within this new theory, Casimir-Lifshitz interactions predicted thin dry ice films in water vapor.
When structural forces are included\,\cite{LuengoMarquez_IzquierdoRuiz_MacDowell2022}, together with Casimir-Lifshitz forces, however, we expect an atomically thin liquid water layer. Notably, referring to recent work on ice premelting by MacDowell and co-workers\,\cite{LuengoMarquez_IzquierdoRuiz_MacDowell2022}, we expect that the outer ice surface can melt via structural forces, leading to a five-layer system with bulk gas hydrate-surface gas hydrate-ice (several tens of nanometers thick)-water (atomically thick)-vapor.

\par Finally, a comment ought to be given on the fundamental physics of the system analyzed here. Throughout our work, we made use of the continuum electrodynamic theory, even for the small distances between dielectric boundary layers. As is well known, the volume force density in such a layer can be expressed as $\mathbf{f}=-\frac{1}{2}\varepsilon_0E^2{\bf \nabla}\varepsilon$, and so the surface force density (pressure) is obtained by integrating the normal component $f_n$ of $\mathbf{f}$ across the layer. This simple theory works well down to quite low distances of the nanoscale. For even smaller distances, the continuum theory has to be abandoned together with some form of statistical mechanics introduced instead. In this context, note the simulation results recently obtained by Serwatka {\it et al.} \cite{serwatka23} on phase transitions in a one-dimensional chain of water molecules. Since water molecules are rotating, it is possible for the system, at low temperatures (about $10\rm\ K$), to perform a phase transition to a ferroelectric phase. The physical reason for the increasing interaction between water molecules with decreasing distances is the rise of quantum fluctuations in the ground state. Analyses in Ref.~\cite{serwatka23} refer primarily to quantum physics. While this area of research is still in its early stages, the results are worth noticing, not only from a quantum standpoint but also from a classical perspective. Behaviors of water molecules at micro or nanoscale distances are complex and extend beyond the scope of the continuum theory.


\begin{acknowledgments}
We thank the Terahertz Physics and Devices Group, Nanchang University for the strong computational facility support.
The authors thank the "ENSEMBLE3 - Centre of Excellence for nanophotonics, advanced materials and novel crystal growth-based technologies" project (GA No. MAB/2020/14) carried out within the International Research Agendas programme of the Foundation for Polish Science co-financed by the European Union under the European Regional Development Fund and the European Union's Horizon 2020 research and innovation programme Teaming for Excellence (GA. No. 857543) for support of this work.
\end{acknowledgments}

\bibliography{drywet}

\end{document}